# Beyond the Turing Threshold: Productive Grammars generate Essentially Undecidable Languages*

Luis M. Augusto

October 21, 2026

Independent Researcher

https://orcid.org/0000-0001-9097-5722

luis.ml.augusto@protonmail.com

**Abstract**

*Emil Post's productive sets are not even semi-computable, let alone computable, being thus essentially incomputable. Accordingly, formal languages whose set of words is a (completely) productive set are essentially undecidable. In this article, I elaborate on Post productivity from the viewpoint of formal language theory: I design formal grammars that emulate the construction of productive sets of natural numbers and are thus beyond Turing-decidability.*

## 1 Introduction

A core construct in formal language theory is the *(extended) Chomsky hierarchy*: It distinguishes formal languages into Turing-decidable and only Turing-recognizable, leaving us essentially ignorant with respect to the uncountably infinitely many languages that are beyond the grasp of the Turing machine (see Fig. 1), which since Turing (1936) remains—and is bound to remain—as the decider of what is or is not computable in the mathematical universe of functions, sets, and relations. This result is well-known in the literature as the *Church-Turing Thesis.*[1]

Regardless of whether one's philosophy of mathematics favors construction over discovery or vice-versa, *productive sets* must be taken under the viewpoint of set construction. These infinite sets, originally constructed in Post (1944), are incomputable; in fact, they are not even semi-computable, being thus essentially incomputable. Thus, one naturally wonders what generation or

*Published as a working paper under the same title in Ω ::=*Journal of Formal Languages*, 2, 35-65; submitted for standard publication in September 2026.

[1]A rather intuitive formulation of this thesis is as follows: *A function that is effectively calculable is a function that can be computed by a Turing machine.* This formulation goes beyond *Church's Thesis*, which merely states that any effectively calculable function is so if and only (abbr.: iff) if it is recursive. Tellingly, the so-far purely hypothetical range of computations beyond the limitations imposed by the Turing machine is called *super-Turing computation* (also: *hypercomputation*).

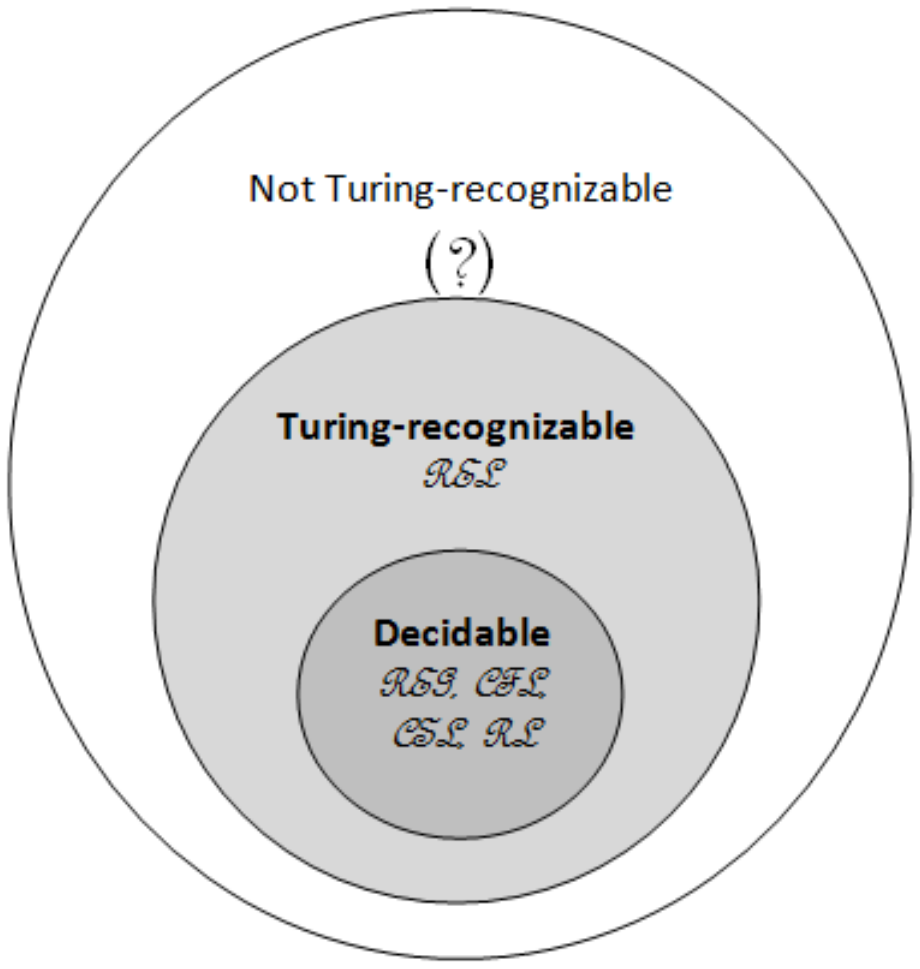

Figure 1: The Chomsky hierarchy and beyond. (Source: Augusto, 2021.)

construction processes yield such sets. If we consider them—as I shall do here—from the viewpoint of the Chomsky hierarchy of formal languages, then productive word sets are wholly outside it, i.e. they are not Turing-recognizable, let alone -decidable, and thus justify our focusing on them for providing an adequate framework to study essentially undecidable languages.[2] From this viewpoint, one might wish to find answers to questions such as: What kind of formal system(s) generate(s) such languages? Is there a class of formal grammars that can be associated with them? I address these questions in this paper. The aim of this study is thus to reduce our ignorance with respect to the uncountably infinitely many languages that are not amenable to the recognition or decision powers of the Turing machine. More specifically, I shall show that *productive grammars*, outside the Chomsky hierarchy and its extensions, generate *productive languages*, i.e. sets of words that behave like productive sets of natural numbers.

Except for Dekker (1955), productive sets are typically studied as the complements of creative sets, and thus the literature on these sets is rather scarce. In effect, creativeness is seen as the one property that "is shared by all the naturally arising unsolvable problems discovered by Church, Turing, and others in the 1930s and after" (Cooper, 2004), but there are uncountably many productive sets without a creative set as their complement, which allows for research on them independently from the larger research on creative sets. Here, I am interested in productive sets per se, namely inasmuch as they provide the adequate framework to approach the questions specified above. I draw mostly on Post (1944) and Dekker (1955), with the odd bits from standard monographs or textbooks, to wit, Cutland (1980), Rogers (1967), Smullyan (1993), and Soare (2016).

The present article is developed within two subjects that are more often than not conflated, to wit, recursiveness and computability. Computability theory is frequently just the name given to recursion theory when we are more interested in practical computing than in the associated mathematical formalisms and structures, but in fact the two are not necessarily one and the same subject. Take the concepts *computable* and *recursive* all too often seen as synonymous to char-acterize a set. This synonymy is however not complete. When defining a computable set in the context of computability theory, we might—and perhaps ought to—take physical resources into consideration.

[2]See Table 1 for the nomenclature used in Figure 1.

**Example 1.** Consider the following set:

$$S = \{p \,|\, p \text{ is a polynomial over } x \text{ with an integral root}\}$$

We ask whether $S$ is computable. Given a polynomial $p$ over $x$ such as $2x^3 + 6x^2 - x + 5$, we ask if $p \in S$. In other words, we want to know if $x = n, n \in \mathbb{Z}$. There is indeed an algorithm to test for this decision problem: We evaluate $p$ for the values $0, 1, -1, 2, -2, 3, ...$ one at a time successively; if for some $n \in \mathbb{Z}$ it is the case that $p = 0$, then $\chi_S(p) = 1$ (cf. Def. 6 below). Nevertheless, we have no guarantee that this algorithm will ever terminate: It will only eventually terminate if indeed $p \in S$; otherwise, it is obvious that it will run forever, because $\mathbb{Z}$ is an infinite set. Thus, $S$ is in fact incomputable, though it is semi-computable.

The physical resource at play in this example is time, but we could consider space instead. When in recursion theory we say that a set is recursively enumerable—seen as a synonym for semi-computable—but not recursive, we say basically the same thing, but we prove it by either showing that there is a recursively enumerable (abbr.: r.e.) set $A$ whose complement is not r.e. or by diagonalization, it being the case that in either case a contradiction is reached; the example given is almost always the set $K = \{x \,|\, x \in Dom(\varphi_x)\}$, and as it will be seen below finite time, or any other physical resource or constraint, plays no role whatsoever in the proof.[3]

It seems to me thus that this disparity in perspective is more than a simple nuance; it shows that Church's concept of *effective calculability* is taken with fundamentally different meanings in the two subjects. More specifically, this concept, originally given in Church (1936), is taken in a practical sense in computability theory and rather intuitively in recursion theory, it being the case that mathematical intuition entails here the notion of *induction*, as argued in Soare (1996, 1999). Importantly, a function need not be recursive to be computable, as the zero function and the projection function show (cf. Def. 3 below). In any case, I shall use the terms *computable* and *recursive* differentially: I shall reserve the latter for functions and the former for sets. To these two terms I shall add a third, to wit, *decidable* to characterize word sets that are computable. In Section 3, I duly justify this addition to a terminology that often requires clarification of its synonyms, precisely because *recursion*, *computability*, and *decidability* are concepts that have emerged and undergone nuanced developments in three subjects that largely overlap.

# 2 Post's Productive Sets

## 2.1 Some Remarks on Functions and Sets from the Viewpoint of Recursion Theory

### 2.1.1 2.1.1. Recursiveness and computability

I begin by recalling some core concepts of recursion theory on functions and sets.

[3]To be sure, the proofs by the Complementation Theorem or by diagonalization assume (often implicitly) Church's Thesis, i.e. a method of effective calculability such as a Turing machine, which in turn assumes an algorithmic procedure. But the onus of the proofs in recursion theory does not fall on the inefficiency of any such procedure, as in Example 1: They show instead by *reductio ad absurdum* that there *cannot* be any algorithm that computes $K$.

**Definition 2.** Let $\mathbb{N} = \{0,1,2,...\} = \omega, |\omega| = \infty = \aleph_0$, be the set of the natural numbers.[4] $\mathbb{N}$ is a *well-ordered set*, as every subset $N \neq \emptyset$ of $\mathbb{N}$ has a least element. An $n$-ary function $f(x_1,...,x_n) = f^n(\overline{x})$ over $\mathbb{N}$ s.t. $f : \mathbb{N}^n \longrightarrow \mathbb{N}$ is said to be *total* if it is defined for every element in its domain, denoted by $Dom(f) = W_f^{(n)}$, and *partial* if it is defined only for a subset of its domain. The range of $f$ is here denoted by $Range(f) = E_f^{(n)}$.

A function $f$ that is partial is denoted by $\varphi$. I shall consider indifferently $\mathbb{N}$ or $\mathbb{Z}^+$.[5] As usually, I write $f(\overline{x})\downarrow$ to denote that the (partial) function $f(\overline{x})$ is defined for all the $x_i$ in $\overline{x}$, and $f(\overline{x})\uparrow$ otherwise. Instead of the subscript "$f$" (or "$\varphi$") we write "$e$", the index of $f$ (or $\varphi$), i.e. the Gödel number of a given program $P$ s.t. $P_e$ computes $f$ (respectively $\varphi$). (In this article, $e$ will be employed to identify an arbitrary program, and thus no actual computation of a Gödel number will be carried out.) Indices other than $e$ will also occur, as any number $n$ is an index of a (partial) function $f(x)$ if $g_n(x) = f(x)$.

**Definition 3.** A function $f$ is said to be *primitive recursive* if it can be generated from the functions $z(x) = 0$, $succ(x) = x+1$, and $U_i^n(\overline{x}) = x_i$—known as zero function, successor function, and projection function, respectively—by zero or more applications of the following schemata known as the composition or substitution schema (1) and the primitive recursive schema (2):

$$(1) \qquad f(\overline{x}) = g(h_1(\overline{x}),...,h_m(\overline{x}))$$

$$(2) \qquad \begin{cases} f(0,\overline{x}) = f(\overline{x}) \\ f(y+1,\overline{x}) = f(y, f(y,\overline{x}), \overline{x}) \end{cases}$$

If

$$(3) \qquad f(\overline{x}) = \mu y R(\overline{x}, y)$$

known as minimalization schema and where $R$ is a relation symbol, is also applied then the function obtained is *partial recursive*.

**Definition 4.** (Church, 1936) A function $f$ is *recursive* iff it is total and *effectively calculable*; it is *partial recursive* (p.r.) iff it is effectively calculable.

The following example introduces two functions that, albeit elementary, will play a major role in the elaboration below on productive sets.

**Example 5.** The functions $f(x) = x$, known as the identity function, and $f_x(x) = x+1$, known as the diagonal function, are recursive. In effect, $f(x) = x+1$ is the successor function in Definition 3, and it is obviously total; $f_x(x)$ is called diagonal function because it is applied only on the elements in the diagonal of an infinite table or matrix $\boldsymbol{A}$ s.t. we have

$$\begin{bmatrix} \underline{a_{0,0}} & a_{0,1} & a_{0,2} & \cdots & a_{0,m} & \cdots \\ a_{1,0} & \underline{a_{1,1}} & a_{1,2} & \cdots & a_{1,m} & \cdots \\ \vdots & \vdots & \vdots & & \vdots & \\ a_{m,0} & a_{m,1} & a_{m,2} & \cdots & \underline{a_{m,m}} & \cdots \\ \vdots & \vdots & \vdots & & \vdots & \end{bmatrix}$$

[4] Recall that $\aleph_0$ is the cardinality of a countably infinite set, i.e. a set whose elements can be put in a 1-1 correspondence with $\mathbb{N}$.

[5] Enumerations will start with 0 or 1, according to convenience.

for every element of a given set $A$. The diagonal function will be denoted by $\delta(x)$. Then, we have $\delta : A \longrightarrow A$ s.t. $\delta(a) \neq a$ for every $a \in A$ s.t. $a$ is $a_{i,i}$, because $\delta(a) = a + 1$. The identity function, denoted by $\iota(x)$ and which is also obviously total, is the special case of schema (2) when $\overline{x} = x$, i.e. $f(0, x) = x$. (Alternatively, we can see the identity function as the projection function $I_i^n(\overline{x}) = x_i$.) The relationship between these two functions is that, as given by $\delta(a) \neq a$, *the diagonal function is never the identity on A; hence, set A is incomputable*, as there is no effective method to enumerate the rows of $\boldsymbol{A}$.

Definition 4 is actually Church's Thesis; in this, effective calculability is associated with an algorithmic procedure, i.e. a function $f$ is effectively calculable iff there is an algorithm that outputs a value given a certain input value. (Note that an algorithm has only a finite number of instructions that are applied in a finite number of steps.)

**Definition 6.** Let there be given a set $A \subseteq \mathbb{N}$. $A$ is *computable* if the *characteristic function* of $x$ with respect to $A$ is effectively calculable:

$$\chi_A(x) = \begin{cases} 1 & \text{if } x \in A \\ 0 & \text{otherwise} \end{cases}$$

$A$ is *semi-computable* if the *semi-characteristic function* of $x$ with respect to $A$ is effectively calculable:

$$\chi_A(x) = \begin{cases} 1 & \text{if } x \in A \\ \uparrow & \text{otherwise} \end{cases}$$

Thus, if $A$ is computable, then there is an effective method that calculates whether $x \in A$ or $x \in \overline{A}$ for all $x \in \mathbb{N}$, $\overline{A}$ denotes the complement set of $A$. If there is an effective method to calculate $x \in A$, but no such method to calculate $x \notin A$, then $A$ is said to be semi-computable.

*Remark* 7. As anticipated in the Introduction, I am henceforth writing *computable* (instead of *recursive*) and *semi-computable* (instead of *recursively enumerable*) for a set $A$ whenever there is an effective method—i.e. an algorithm—to calculate the (semi-)characteristic function for $A$. I leave the above terminology for functions unchanged, because no function is computable (i.e. effectively calculable) if it is not a primitive recursive function or a composition of primitive recursive functions.

**Definition 8.** A set $A$ that satisfies the condition

$$[(x \in A) \wedge (\varphi_x = \varphi_y)] \Longrightarrow (y \in A)$$

for all $x, y \in \mathbb{N}$ and $\varphi$ a p.r. (unary) function, is called an *index set*.

*Remark* 9. The reader should keep in mind the case when $\varphi_x = \varphi_{f(x)}$, where $f(x) = \iota(x)$ is the identity function. Clearly, if $x \in A$ and $A$ is an index set, then $f(x) \in A$.

**Theorem 10.** *(Rice's theorem) The problem* $\varphi_x \overset{?}{\in} A$ *is undecidable if* $A \neq \emptyset$ *or* $A \neq \mathbb{N}$.

Intuitively put, Rice's theorem formulates the result that the only computable index sets are $\emptyset$ and $\mathbb{N}$. These two index sets are said to be *trivial*, because $\emptyset$ contains no index of any p.r. function $\varphi$ and $\mathbb{N}$ contains all its indices. The meaning of this result is that any non-trivial property of p.r. functions is *undecidable*, a term that will be properly defined below; because the set of all p.r.

(unary) functions is semi-computable this means that every non-trivial property of semi-computable sets is undecidable.

The following definitions and results draw on Post (1944). Effective calculability and enumeration are tightly associated and indeed we have:

**Definition 11.** A set $A \subseteq \mathbb{N}$ is semi-computable if its members can be *enumerated* as

$$A = \{f(0), f(1), f(2), ...\} = Range(f) = E_f = E_e$$

or, equivalently,

$$A = \{0, 1, 2, ...\} = Dom(f) = W_f = W_e$$

where $f$ is recursive and $e$ is the index for $A$.

Recall that "semi-computable" is taken here as a synonym for "recursively enumerable" when speaking of sets; this explains why "computably enumerable," a term also often found in the literature, is a synonym for "semi-computable." The fact that a set $A$ is semi-computable does not mean that it is computable.

**Example 12.** Let us consider the set $K = \{e \mid e \in W_e\}$, known as *Gödel's diagonal set*, and called in Post (1944) *the complete set*, because every semi-computable set $W_e$ is computable in $K$. $K$ is semi-computable. In effect, $K$ is the domain of the p.r. function

$$\psi(x) = \begin{cases} x & \text{if } \varphi_x(x)\downarrow \\ \uparrow & \text{otherwise} \end{cases}$$

which is effectively calculable by applying an algorithm—say, a program $P_x$—to input $x$. But $K$ is incomputable. If $K$ had an effectively calculable characteristic function, then the following function would be effectively calculable:

$$f(x) = \begin{cases} \varphi_x(x) + 1 & \text{if } x \in K \\ 0 & \text{if } x \notin K \end{cases}.$$

Then, $f(x) \neq \varphi_x$ for every $x$, which shows that $f$ cannot be recursive. Indeed, let $f(x) = \varphi_x(x)$; then, we have $\varphi_x(x) = \varphi_x(x) + 1$, and $0 = 1$, which is impossible. Hence, $K$ is semi-computable but incomputable.

Besides by a diagonalization argument, as in Example 12, we can show that a semi-computable set fails to be computable by Post's Complementation Theorem.

**Theorem 13.** *(Complementation Theorem) A semi-computable set $A \subseteq \mathbb{N}$ is computable iff its complement $\overline{A}$ is also semi-computable*

*Proof.* ($\Rightarrow$) If $\chi_A(n) = 1$, then put $n$ in $A$. Otherwise, put $n$ in $\overline{A}$. Hence, both $A$ and $\overline{A}$ are semi-computable.

($\Leftarrow$) Let $A, \overline{A}$ be semi-computable sets s.t. $A = \{n_1, n_2, ...\}$ and $\overline{A} = \{m_1, m_2, ...\}$. Given some $n \in \mathbb{N}$, generate successively $n_1, m_1, n_2, m_2, ...$ comparing with $n$. Obviously, the end result of our generation and comparison must be that either $n \in A$ or $n \in \overline{A}$, so in a finite number of steps we shall find the $n_i$ or the $m_i$ that is identical with $n$. This is thus an effective method to determine the membership of any $n$ to either $A$ or $\overline{A}$. Hence, $A$ is computable. □

The concept of *generated set* introduced in the $\Leftarrow$-direction of Post's proof of Theorem 13 will play an important role in this work. We have the following important result (which here is a Proposition, but is just part of the main text of Post's paper):

**Proposition 14.** *Every generated set of positive integers is semi-computable.*

*Proof.* This can be restated in two statements, to wit, (a) every generated set is effectively calculable, and (b) every effectively calculable set of positive integers is semi-computable. With respect to (a), "effectively calculable" just means the same as "effectively enumerable," i.e. all one need have is an effective process to enumerate elements into a set as belonging to it; any recursive function $f$ will be such a process. We have then the set $\{f(i)\}_{i=0}^{k \text{ or } \omega}$, which is semi-computable by Def. 5, and we are done with (b). □

With respect to the set $A = \{f(i)\}_{i=0}^{k \text{ or } \omega}$ of the proof above, we have the following remark:

*Remark* 15. The set $\{f(i)\}_{i=0}^{k}$, $k$ is finite, is computable. The set $\{f(i)\}_{i=0}^{\omega}$ is computable iff it admits of a recursive enumeration in order of magnitude, and non-computable otherwise. In other words, every finite set is computable, and an infinite semi-computable set is computable iff it is the range of an increasing function $f$.[6] Importantly, this also tells us that semi-computable sets that are incomputable are always infinite.

**Theorem 16.** *Every infinite semi-computable set $A$ contains an infinite computable subset $A'$.*

*Proof.* Generate in order the elements of the recursive enumeration $\alpha = n_1, n_2, ...$ of an infinite set $A$ and let $m_1 = n_1$. As for $m_2$, let $m_2 = n_{i_2}$, i.e. the first $n_i$ greater than $n_1$, and so on successively; because $\alpha$ is a recursive enumeration we are assured that for every $n_i$ there is an $n_j$ later in $\alpha$ s.t. $n_j > n_i$. We thus generate the set $A' = \{m_1, m_2, ...\}$ without repetitions in order of magnitude. Clearly, $A'$ is a subset of $A$ that is infinite and computable. □

**Theorem 17.** *There is a semi-computable set $A$ of positive integers that is incomputable.*

*Proof.* By Post's Complementation Theorem, $\overline{A}$ is not semi-computable. We show this by the diagonalization method. Let $U = \{A_i \mid i = 1, 2, ...; A_i \text{ is s.c.}\}$. "s.c." abbreviates "semi-computable", be a set of generated elements; then, $A$ is semi-computable by Proposition 14. More precisely, $U$ is the set of all the semi-computable sets. A positive integer $x$ thus is, or is not, in $U$ according as it is, or is not, in the $i$-th set $A_i$. So, if $x$ is not in $U$, then it must be in $\overline{U}$; but then it is not in the $i$-th semi-computable set in $U$, and $\overline{U}$ differs from each semi-computable $A_i$ in the presence or absence of at least one positive integer. Hence, $\overline{U}$ is not semi-computable. Now, make $U = A$, and QED. □

**Example 18.** The set $K = \{x \mid x \in W_x\}$ is semi-computable but incomputable. This was shown in Example 12 by an application of diagonalization (i.e. by employing the diagonal function). This also means that $\overline{K} = \{x \mid x \notin W_x\}$ is not semi-computable. Assume otherwise that $\overline{K}$ is also semi-computable. Then, for some $e \in \mathbb{N}$ we must have $W_e = \overline{K}$, thus giving

$$x \in W_e \Leftrightarrow x \in \overline{K} \Leftrightarrow x \notin K \Leftrightarrow x \notin W_x$$

We get a contradiction when making $x = e$, so we conclude that $\overline{K}$ is not semi-computable.

[6] Recall that a function $f$ over $\mathbb{N}$ is *increasing*, or *order-preserving*, if $f(x) < f(y)$ whenever $x < y$ for $x, y \in \mathbb{N}$.

**Corollary 19.** *An incomputable set that is not semi-computable is essentially incomputable.*

*Proof.* (Informal) Example 18 above shows that the attempt to compute $\overline{K}$ leads to a contradiction. In effect, note the paradoxical character of $\overline{K}$, kindred with Russell's paradox: As it is possible to have $W_n = \{n\}$, then we have $\overline{K} = \{n \mid n \notin \{n\}\}$. Hence, $\overline{K}$ is essentially incomputable. □

### 2.1.2 2.1.2. Infinite sets

Above, a few references, explicit or implicit, were made with respect to infinite sets. Infinity plays a central role in this work, namely in relation to the following result, which is directly connected with the proof of Corollary 19 immediately above:

**Proposition 20.** *Every incomputable set is always infinite.*

For the present study, it suffices to grasp the concept of *natural order*, i.e. the well-ordering of the natural numbers, denoted by the symbol "$<$". I assume that the reader is acquainted with the principle of induction. With this principle in hand, one way to define infinity mathematically is by the following statement that elaborates formally on the fact that *there exists an infinite set*:[7]

*Infinity Axiom:* There exists at least a set $I$ s.t. (i) $\emptyset \in I$, and (ii) if $x \in I$, then also $\{x\} \in I$. Formally:

$$\exists I\, [\emptyset \in I \wedge \forall x\, (x \in I \Rightarrow (x \cup \{x\}) \in I)]$$

**Definition 21.** Let $I$ be a set. We say that $I$ is an *inductive set*, and we write $\breve{I}$, iff $I$ is a witness to the Infinity Axiom, i.e.:

$$\breve{I} \quad \text{iff} \quad [\emptyset \in I \wedge \forall x\, (x \in I \Rightarrow (x \cup \{x\}) \in I)]$$

We can then account constructively for the existence of the set $\mathbb{N}$ in the following way:

**Proposition 22.** *There is a unique set $I$ s.t. $I \subseteq J$ for every inductive set $J$.*

*Proof.* (Sketch) Starting from (1) $\breve{I} \cap \breve{J}$ is an inductive set, and (2) for $\breve{K}$, we let

$$\mathcal{D} = \left\{ J \in 2^{\breve{K}} \mid J \text{ is } \breve{J} \right\}$$

and

$$I = \left\{ x \in \breve{K} \mid \forall J \in \mathcal{D} : x \in J \right\}$$

we show that (a) $I$ is an inductive set, (b) for every $\breve{J}$ we have $\breve{I} \subseteq \breve{J}$, and (c) if $I'$ is an inductive set and for every inductive set $J$ we have $\breve{I}' \subseteq \breve{J}$, then $\breve{I}' = \breve{I}$. □

More intuitively, if we let $n = \{0, 1, 2, ..., n-1\}$ in the sense that the set $n$ represents the natural number $n$, then

$$0 = \emptyset, \quad 1 = \{0\}, \quad 2 = \{0, 1\}, \quad 3 = \{0, 1, 2\}, \quad ...$$

This can be expressed in the intuitive principle that *each natural number n should have n elements.* Then, by applying induction we have

$$0 + 1 = \emptyset \cup \{0\} = 1,$$

[7] This is one of the seven axioms of the axiomatization of set theory known as *Zermelo-Fraenkel set theory with the Choice Axiom* (*ZFC* for short). See Augusto (2020) for a brief discussion and references to the standard literature.

$$1 + 1 = \{0\} \cup \{1\} = 2$$

$$2 + 1 = \{0, 1\} \cup \{2\} = 3$$

$$\vdots$$

$$n + 1 = n \cup \{n\}$$

for every $n \in \mathbb{N}$, and $\mathbb{N}$ is the smallest inductive set in the sense that

$$\mathbb{N} = \bigcap \mathcal{D} = \bigcap I.$$

The formulation above of the Infinity Axiom implies that $\mathbb{N} \subseteq \breve{I}$, and if additionally we know that $\mathbb{N} \supseteq \breve{I}$, then we can formulate it as the *induction property of* $\mathbb{N}$:

$$\exists I \left[\emptyset \in I \wedge \forall n \left(n \in I \Rightarrow n + 1 \in I\right)\right] \Rightarrow \left(\mathbb{N} = \breve{I}\right)$$

*Remark* 23. It is now obvious that we can rephrase the Infinity Axiom as: *There exists a reflexive set.* In effect, if we consider $x \in \breve{I}$ as a first element, we have the set

$$\breve{I} = \{x, \{x\}, \{x, \{x\}\}, \{x, \{x\}, \{x, \{x\}\}\}, ...\}$$

where the notion of *ordering* is evident: We have

$$x < \{x\} < \{x, \{x\}\} < ...$$

s.t. $\{x\}$ is the *successor* of $x$, etc. Denoting this relationship by $Succ\left(n\right)$ for $n \in \mathbb{N}$, we have:

$$Succ\left(n\right) = n + 1 = n \cup \{n\}$$

Let us now start with the first element $\emptyset \in \breve{I}$; then we have

$$\mathbb{N} = \{\emptyset, \{\emptyset\}, \{\emptyset, \{\emptyset\}\}, \{\emptyset, \{\emptyset\}, \{\emptyset, \{\emptyset\}\}\}, ...\}$$

By correlating every $x \in \breve{I}$ with $\{x\} \in \breve{I}$, we obtain a one-to-one mapping of $\breve{I}$ onto a proper subset of $2^{\breve{I}} - \emptyset$.

There are three notations used to denote infinity, to wit, $\omega$, $\infty$, and $\mathbb{N}$. They are not equivalent in the present work, though. As mentioned above, we have typically

$$\mathbb{N} = \{0, 1, 2, 3, ...\} = \omega$$

but we prefer to settle things so that $\mathbb{N}$ denotes the infinite set of natural numbers and

$$|\mathbb{N}| = \infty$$

i.e. $\infty$ denotes the cardinality of $\mathbb{N}$, also often denoted by $\aleph_0$, without regard to order.

*Remark* 24. But the two notations $\omega$ and $\infty$ may coincide in the notion of *tending to infinity.* Recall that an infinite sequence is defined as

$$(a_n) = \{a_n \mid n \in \omega\}$$

where $a_n$ denotes the $n$-th term of infinitely many terms belonging to $\omega = \mathbb{N}$ or $\omega = \mathbb{Z}^+$. Let now $f(n)$ be the identity function $\iota(n)$, i.e. $f(0) = 0$, $f(1) = 1$, ...; then we have

$$\lim_{n \to \infty} f(n) = \infty$$

where $n \to \infty$ and $f(n) = \infty$ just is the conventional notation of mathematical analysis, but because $n \in \omega$ we may write instead $f(n) = \omega$. In particular, if we consider $n \in \omega$ as a superscript s.t. the $n$-th term coincides with the $n$-th potency in a sequence s.t. $a^n = \overbrace{a...a}^{n \text{ times}}$, then we have for an unbounded sequential function $f(a^n)$:

$$\lim_{n \to \infty} f(a^n) = a^{\infty} = a^{\omega}$$

**Example 25.** Let $a = 011$ and $n \in \mathbb{Z}^+$ s.t. $a_n = (011)^n$. Then, $a^1 = 011$, $a^2 = 011011$, ..., $a^n = \overbrace{011}^{n \text{ times}}$, and

$$\lim_{n \to \infty} (011)^n = (011)^{\infty} = (011)^{\omega}.$$

Note that $(011)^{\omega}$ provides us with a means to *represent finitely* the infinite string 011011011... corresponding to

$$\underbrace{\underbrace{\underbrace{\underbrace{011}_{1}011}_{2}011}_{3}\ldots}_{\vdots}$$

a rather cumbersome representation. That is to say that we employ the notation $\omega$ to denote an infinite sequence *ordered* by $\mathbb{N}$ (or any infinite subset of $\mathbb{N}$); in effect, $\omega$ is the standard symbol for the first infinite ordinal.[8] Thus, I shall employ the notation $\omega$ also for infinite cardinality if an infinite sequence is implied.

*Remark* 26. One way to explain intuitively the difference between the two notations for infinity, $\omega$ and $\infty$, is as follows: Let there be given some infinite set $A$; if we look upon this set *statically* as a constructed set, then we write $|A| = \infty$, but if our view on it is *dynamic*, in the sense that we are interested in how it is constructed, i.e. how its elements get to belong to it *one by one*, or sequentially, then we write $|A| = \omega$. An important aspect of this dynamic viewpoint on infinity is that at any one time of this construction we can stop and see the obtained objects (e.g., strings of symbols) as constituting a finite set. Take for example the number $\pi = 3.14159...$; we denote the infinite character of its decimal expansion by $|\pi| = \infty$, but we write $|\pi|_5 = \omega$ to denote that there are infinitely many 5's in the decimal expansion of $\pi$, the first 5 in the 4th decimal position, etc. We can thus call these two perspectives respectively *static infinity* and *dynamic infinity.*

[8] The natural numbers less than $\omega$ are also called *finite ordinals*, because a set $A$ is said to be finite if there is a one-to-one mapping of $A$ onto some $n \in \mathbb{N}$. By starting with $0 = \emptyset$, $1 = \{0\}$, ..., and using the operation $n + 1 = n \cup \{n\}$ we reach the next ordinal after all the ordinal numbers, i.e. $\omega = \{0, 1, 2, ...\}$. Thus, $\omega$ is the first *infinite ordinal.* By applying now the operation $\omega + 1 = \omega \cup \{\omega\} = \{0, 1, 2, ..., \omega\}$, we obtain the infinite sequence of ordinals $\omega + 1, \omega + 2, \omega + 3, ...$, which in turn is followed by infinitely more infinite ordinals $\omega + \omega, \omega + \omega + 1, ...$, etc.

## 2.2 Productive Sets

Recall now from Rice's theorem (cf. Theorem 10) that the set of the natural numbers $\mathbb{N}$, just like $\emptyset$, is a trivial index set. Above, $\mathbb{N}$ was approached dynamically, i.e. it was shown how each natural number $m \in \mathbb{N}$ iff $m = n + 1$ for $n \in \mathbb{N}$, i.e. there is a recursive function $succ : \mathbb{N} \longrightarrow \mathbb{N}$ s.t. $succ\,(n) = n + 1 = m$. It follows that $\mathbb{N}$ is effectively enumerable; in fact, $\mathbb{N}$ is computable, as for every $n, m \in \mathbb{N}$ s.t. $succ\,(n) = m$, $m > n$. In effect, let $W_0 = \emptyset$, $W_1 = \{0\}$, $W_2 = \{0, 1\}$, ..., and apply the identity function to $x_i = f\,(i)$, which is obviously order-preserving, as $f\,(i) < f\,(i+1)$ for $x_i < x_{i+1}$. We then have:

$$W_0 = \emptyset$$
$$W_1 = (W_0 \cup \{f\,(0)\}) = (\emptyset \cup \{0\}) = 1, \text{ and } f\,(1) \in W_1$$
$$W_2 = (W_1 \cup \{f\,(1)\}) = (1 \cup \{1\}) = 2, \text{ and } f\,(2) \in W_2$$
$$\vdots$$
$$W_n = \left( \underbrace{\{0, 1, ..., n-1\}}_{W_{n-1}} \cup \{f\,(n-1)\} \right) = (n - 1 \cup \{n-1\}) = n, \text{ and } f\,(n) \in W_n$$
$$W_{n+1} = \left( \underbrace{\{0, 1, ..., n-1, n\}}_{W_n} \cup \{f\,(n)\} \right) = (n \cup \{n\}) = n + 1, \text{ and } f\,(n+1) \in W_{n+1}$$

so that we have:

$$\mathbb{N} = \bigcup_{n=0}^{\omega} W_n = \bigcup_{n=0}^{\omega} f\,(n)$$

What happens now if we construct a set, say $A \subseteq \mathbb{N}$, that is obviously inductive but s.t. for every $x_i \in A$ it is the case that $f\,(i) \notin (W_i \subseteq A)$, i.e. there is a number $x_i$ in $A$ not in $W_i \subseteq A$? We then have $f\,(i) \in (A - W_i)$ and $f\,(i) \neq x_j$ for $j = 0, 1, ..., i-1$, and $A \supseteq \{x_0, x_1, ..., x_{i-1}, f\,(i)\}$ is incomputable, though it is inductive; in particular, $A$ clearly has an infinite semi-computable subset $A' = \{f\,(0), ..., f\,(n), ...\}$. Such a set $A$ is said to be *productive*, and its relationship with Rice's theorem can be formulated as follows: If $A \neq \emptyset$ or $A \neq \mathbb{N}$, then $A$ is either productive or the complement of a productive set. This result is directly mirrored in complexity theory in the following way: The productive sets, as well as their (semi-computable) complements, constitute essentially undecidable problems. In effect, productive sets are constructed in analogy with the well-known diagonal method to prove the incomputability of the interval $(0, 1] \in \mathbb{R}$; in other words, there is no effective method (e.g., a Turing machine) to enumerate the elements of a productive set. Hence, and paradigmatically, the complete theory of the natural numbers $\mathcal{N}$ is productive and the system of Peano arithmetic is creative.[9]

[9]These correspond to the sets

$$\mathcal{N} = \{\psi_e \,|\, \psi \in L\,(\mathbb{N}) \text{ and } \langle \mathbb{N}, +, *, 0, 1, \leq \rangle \models \psi\}$$

**Definition 27.** A set $A \subseteq \mathbb{N}$ is called *productive* if there is a recursive function $f(x)$ s.t. for any $n \in \mathbb{N}$,

$$W_n \subseteq A \quad \Rightarrow \quad f(n) \in (A - W_n)$$

We say that $f(x)$ is a *productive function* for $A$ and $A$ is productive relative to $f(x)$.

*Remark* 28. This definition entails that we have

$$|A| >^1 \left| \bigcup_n W_n \subseteq A \right|$$

where $>^1$ denotes "greater than by (at least) one element", i.e. no semi-computable subset $W_n$ of $A$ is $A$, or, equivalently, there is always one element $n$ in $A$ that does not belong to any of its subsets $W_n$. Clearly, we have $|A| = \omega$ (cf. Remark 26).

These results will be formally proven below.

**Definition 29.** A set $B$ of positive integers whose complement $\overline{B}$ is a productive set is called *co-productive*. A productive function $f(x)$ for $\overline{B}$ is a *co-productive* function for $B$.

**Example 30.** The prototypical example of a productive set is $\overline{K} = \{x \mid x \notin W_x\}$; thus $K$ is a co-productive set with respect to $\overline{K}$ (cf. Example 18). The productive function for $\overline{K}$ (the co-productive function for $K$) is the identity function $\iota(x) = x$. Examples of other productive sets, where "other" is to be taken in the sense that $\overline{K}$ is the prototypical productive set, are:[10]

$$Tot = \{x \mid \varphi_x \text{ is total}\} = \{x \mid W_x = \omega\}$$
$$Fin = \{x \mid |W_x| < |\omega|\}$$

This example shows that for a productive set $A$ we can effectively find a natural number $n$ that is in $A$ but not in $W_n$. Such a number is called a *witness* (that $A \neq W_n$, $\forall n$). Importantly, Remark 26 also entails the following results:

**Lemma 31.** *For $g(x)$ a recursive function, there is a recursive function $k(x)$ s.t. for every $n \in \mathbb{N}$ we have*

$$W_{k(n)} = W_n \cup \{g(n)\}$$

*and the set $A$ s.t. $g(x) \in A$ is productive relative to $k(x)$.*

*Proof.* Define

$$\varphi_{k(n)}(y) = \begin{cases} 1 & \text{if } y \in W_n \text{ or } y = g(n) \\ \uparrow & \text{otherwise} \end{cases} .$$

Then, we have

$$W_{k(n)} = W_n \cup \{y\}.$$

Let now $y \in A$; it is readily seen that $A$ is productive relative to $k(n)$. □

---

and

$$\mathcal{PA} = \{\psi_e \mid \psi \in L(\mathbb{N}) \text{ and } \boldsymbol{PA} \models \psi\}$$

where $L(\mathbb{N})$ is first-order predicate language with identity, $\boldsymbol{PA}$ denotes the axiomatization of Peano arithmetic, and $e$ is the Gödel number of $\psi$. See Augusto (2020) for the symbol "$\models$", which denotes logical consequence.

[10] "Prototypical," in turn, is taken in the sense that a set $A$ is productive iff $\overline{K} \leq_m A$, where "$\leq_m$" denotes many-one reducibility. (The reader is referred to any of the cited texts on computability for elaborations on the diverse types of reducibility. In particular, Odifreddi, 1992, covers them rather exhaustively.)

**Theorem 32.** *If $A \subseteq \mathbb{N}$ is productive, then $A$ has an infinite semi-computable subset.*

*Proof.* Let $A$ be productive with productive function $f(i)$, $i \in \mathbb{N}$ abbreviates the index $e_i$, and let $y_0 = f(0)$. Because $A$ is productive, we have $y_0 \in (A - (W_0 = \emptyset))$. Obtain now $y_1 = f(1)$ s.t. $y_1 \in \left(A - \left(\underbrace{W_0 \cup \{y_0\}}_{W_1}\right)\right)$, then an element $y_2 = f(2)$ s.t. $y_2 \in \left(A - \left(\underbrace{\overbrace{\{y_0\}}^{W_1} \cup \{y_1\}}_{W_2}\right)\right)$, etc. The set $W_{n+1} = \{y_0, y_1, ..., y_n\}$ is a generated set, and hence it is semi-computable by Proposition 14, but $y_{n+1} \in \left(A - \left(\underbrace{\overbrace{\{y_i\}_{i=0}^{n-1}}^{W_n} \cup \{y_n\}}_{W_{n+1}}\right)\right)$, so $y_{n+1} \neq y_0, y_1, ..., y_n$ for every natural number $n \geq 0$. Hence, the set $\{y_0, y_1, ...\} \subsetneq A$ is an infinite semi-computable set. □

The above results allow us to redefine a productive set in the following way:

**Definition 33.** A set $A \subseteq \mathbb{N}$ is productive iff there is a recursive function $k(x)$ s.t. for all $n \in \mathbb{N}$,

$$W_n \subseteq A \quad \Rightarrow \quad \begin{cases} (1) \quad W_{k(n)} \subseteq (A - W_n) \\ (2) \quad \left|W_{k(n)}\right| = |\omega| \end{cases} .$$

**Corollary 34.** *If $A \subseteq \mathbb{N}$ is productive, then $A$ is not semi-computable.*

*Proof.* If $A$ were semi-computable, there could be no positive integer $y_{n+1}$ in $A$ not in the semi-computable subset $W_{n+1} \subseteq A$. But $y_{n+1} \in (A - W_{n+1})$ where $W_{n+1} = \left(\{y_i\}_{i=0}^{n-1} \cup \{y_n\}\right)$, and consequently $A$ is not semi-computable. In effect, we would have it that there would be a natural number $n + 1 \in [(A - (W_n = A)) = \emptyset]$, an absurdity. □

**Definition 35.** A set $A \subseteq \mathbb{N}$ is *completely productive* (c.p.) if there is a recursive function $g(x)$ s.t. for every $n \in \mathbb{N}$ we have

$$g(n) \in A \Leftrightarrow g(n) \notin W_n$$

or equivalently

$$g(n) \in [(A - W_n) \cup (W_n - A)] .$$

The function $g(x)$ is said to be a *c.p. function* of $A$ and $A$ is said to be *c.p. relative to* $g(x)$.

**Example 36.** $\overline{K}$ is c.p. with $\iota(x)$ as c.p. function. In effect, by the definition of the set $K$, $x \in W_x \Rightarrow x \notin \overline{K}$ and $x \notin W_x \Rightarrow x \in \overline{K}$.

**Proposition 37.** *Every c.p. set $A$ is productive.*

*Proof.* $W_n \subseteq A$ implies that $[(A - W_n) \cup (W_n - A)] = (A - W_n)$. □

(It is not known whether there is a productive set that is not c.p.)

*Remark* 38. Obviously, $[(A - W_n) \cup (W_n - A)] \neq \emptyset$, and so again we have $A \neq W_n$, and $n$ is a witness for the fact that $A$ is c.p.

**Definition 39.** Let $A \subseteq \mathbb{N}$ be a productive set and $\Psi$ an effective procedure to find for every semi-computable subset $W_n$ of $A$ a witness to the fact that $W_n \subset A$. The set of all witnesses of $A$, called the *productive center of $A$ with respect to* $\Psi$, is denoted by $\Pi_\Psi(A)$.

Clearly, $Dom\,(A) := \{n \mid W_n \subseteq A\}$.

**Definition 40.** Let $A \subseteq \mathbb{N}$ be productive relative to $f\,(x)$. Then, the set $\Pi\,(A, f) := f\,(Dom\,(A))$ is called the *productive center of A relative to* $f\,(x)$. The subset $\Pi$ of $A$ is called a productive center of $A$ if $\Pi = \Pi\,(A, f)$ for some productive function $f\,(x)$ of $A$.

**Proposition 41.** *If $\Pi_0$ is a productive center of the productive set A, there is a productive center $\Pi_1$ of A s.t. $\Pi_1 \subseteq \Pi_0$ and $|\Pi_0 - \Pi_1| = \omega$.*

*Proof.* Let $f_0\,(x)$ be a productive function for $A$ s.t. we have $\Pi_0 = \Pi\,(A, f_0)$. Let now $f_1\,(n) := f_0\,(k\,(n))$ for $k\,(n)$ a recursive function s.t. $W_{k(n)} = W_n \cup B$, where $B \subseteq \Pi_0$. Let now $\Pi_1 = (A, f_1)$. Then, it is easy to see that $f_1\,(n)$ is a productive function of $A$ and $\Pi_1 \subseteq \Pi_0$. Hence, $B \subseteq (\Pi_0 - \Pi_1)$ is an infinite semi-computable subset of $A$ and $|\Pi_0 - \Pi_1| = \omega$. □

The following statements are left without proof. The objective is to give an idea of the *infinite productivity* of productive sets.[11]

**Corollary 42.** *Every productive set A has a productive center $\Pi_1$ s.t. $|A - \Pi_1| = |\omega|$.*

**Theorem 43.** *Every productive set has exactly countably many productive centers and exactly countably many productive functions.*

**Theorem 44.** *Let A be productive relative to $f\,(x)$ and let $\Pi\,(A, f) \subseteq B \subseteq A$. Then, B is productive relative to $f\,(x)$ and $\Pi\,(B, f) \subseteq \Pi\,(A, f)$.*

**Corollary 45.** *If A is productive relative to $f\,(x)$, then $\Pi\,(A, f)$ is also productive relative to $f\,(x)$.*

**Corollary 46.** *Every productive set induces exactly k productive sets.*

# 3 Productive Grammars and Essentially Undecidable Languages

## 3.1 Review of terminology and notation

In this Section, I recall the basic aspects of formal language theory that are relevant for the main subject of this paper. For convenience, I divide it into two smaller Sections, one recalling the core aspects of formal language theory and another reviewing Turing-decidability.

### 3.1.1 3.1.1. Basics of formal language theory

**Definition 47.** Let $\Sigma = \{a_1, a_2, ..., a_n\}$ be a set of *symbols* (also: *letters*) called an *alphabet*. If $a_1$ precedes $a_2$, etc. up to $a_n$, a property denoted by $a_1 \prec a_2 \prec ... \prec a_n$, then $\Sigma$ is said to be in *lexicographic order*. A *word* over the set $\Sigma$ is a finite sequential factorization

$$w = a_{i_1} a_{i_2} ... a_{i_k}$$

$i \in \{1, 2, ..., n\}$, of symbols over $\Sigma$ s.t. $k \in \mathbb{N}$ is the length of $w$, a property denoted by $\ell\,(w) = k$. The intervals $(a_{i_1}, a_{i_k}]$, $[a_{i_1}, a_{i_k})$, and $(a_{i_1}, a_{i_k})$ of a word $w$ are called *subwords*, respectively.

[11]The proofs can be found in Dekker (1955).

1. The substring $w' = [a_{i_1}, a_{i_k})$ is called a *(word) prefix.*
2. The substring $w' = (a_{i_1}, a_{i_k}]$ is called a *(word) suffix.*
3. The substring $w' = (a_{i_1}, a_{i_k})$ is called a *(word) infix.*

Examples of alphabets in lexicographic order are the Roman alphabet $\{a, b, ..., z\}$ or subsets thereof and the alphabet $\{0, 1\}$. From the above definition, we have it that every prefix and every suffix is an infix. By comparison to $w = a_{i_1} a_{i_2} ... a_{i_k}$, the above subwords are said to be a *proper suffix*, a *proper prefix*, and a *proper infix*, respectively, as any word can be seen as a suffix, a prefix, or an infix of itself. We denote an arbitrary (sub)word over an alphabet $\Sigma$ by the final letters of the Roman alphabet $u, ..., z$.

**Definition 48.** A *word language* $L$ over $\Sigma$ is a set of words $w$ over $\Sigma$.

Henceforth, I write only "language(s)," as I do not discuss any formal languages other than word languages.

**Definition 49.** A word $w$ s.t. $\ell(w) = 0$ is called the *empty word*, denoted by $\epsilon$ (or $\lambda$). Then, $\Sigma^*$ denotes the set of all words, including the empty word, over $\Sigma$. A language $L$ over $\Sigma$ just is a subset of $\Sigma^*$, i.e. $L \subseteq \Sigma^*$ (and any subset of $\Sigma^*$, including $\emptyset$, is a language). More formally, we define a language $L$ intensionally as:

$$L = \{w \in \Sigma^* \,|\, w \text{ has property } P\}$$

This definition gives us the *complement* of a language $L$ as:

$$\overline{L} = \{w \in \Sigma^* \,|\, w \text{ hasn't property } P\}$$

If a language $L \neq \emptyset$ over $\Sigma$ does not include the empty word, then we write $\Sigma^+$, i.e. $\Sigma^+ = (\Sigma^* - \{\epsilon\})$. Note that $\Sigma^+ = \bigcup_{i=1} \Sigma^i$, where $i \in \mathbb{N}$ denotes the length of the words that are elements of $\Sigma^i$. As $\Sigma^0 = \{\epsilon\}$, we have $\Sigma^* = \bigcup_{i=0} \Sigma^i$.

**Example 50.** Consider the language $L = \{ab, aab, aaab, ..., a^i b, ...\}$. This extensional definition corresponds to the intensional definition:

$$L = \left\{a^n b \in \{a, b\}^{n+1} \,|\, n \geq 1\right\}$$

The complement of this language is constituted by words like $b$, $a^n$, $a^n b^m$ for $m > 1$, etc.

Note in the above definition that $\{a, b\}^{n+1}$ denotes that the words built over the alphabet $\{a, b\}$ are of length $n + 1$ for $n \geq 1$; this is simply a specification of $\{a, b\}^+$. It should be obvious that if $\Sigma = \emptyset$, then $\Sigma^* = \Sigma^0$, but if $\Sigma \neq \emptyset$, then $\Sigma^*$ is infinite, and because formal languages $L$ over $\Sigma$ just are subsets of $\Sigma^*$, they are in infinite number, too. However, a language $L$ over an alphabet $\Sigma$ can be finite or infinite.

**Example 51.** The languages $L = \emptyset$, $L = \{\epsilon\}$, and $L = \{w \in \Sigma^+ \,|\, w \in \textit{Hamlet}\}$ (where *Hamlet* denotes the set of all the words in Shakespeare's play identically entitled) are finite. The language of Example 50 above is infinite.

A language can also be infinite if its words are infinite words. This requires a disambiguation of the expression "infinite language."

**Definition 52.** Let $(w||_j)^\omega$ denote the infinite iteration of the $j$-th letters of the finite word $w$ over an alphabet $\Sigma^+$. Then, $w \in \Sigma^\omega$, where $\Sigma^\omega = \bigcup_{i\geq 1}^{\omega} \Sigma^i$, is called a *periodic $\omega$-word* and $L^\omega = L \subseteq \Sigma^\omega = \bigcup_{i=1}^{\omega} L^i$ is a *periodic $\omega$-language.* A periodic $\omega$-word $w$ s.t. $w||_j$ is a suffix is called an *ultimately periodic $\omega$-word.*

**Example 53.** Given $\Sigma = \{a, b\}$, any word $w \in \{a, b\}^\omega$ in which there are infinitely many iterations of $w||_j$ (e.g., $a^*b^\omega$, $a^\omega b$, $ba^\omega$, $(ab)^\omega$, $ab^\omega a$) is an infinite word over $\Sigma$ and thus constitutes a language $L^\omega$. For instance, the language $L = \{w \in \{a, b\}^\omega \mid w = ab^\omega\}$ is an $\omega$-language, namely an ultimately periodic $\omega$-language. In effect, we have:

$$L = \left\{ab, abb, ..., ab^i, ...\right\}$$

I shall abbreviate "(ultimately) periodic $\omega$-word" and "(ultimately) periodic $\omega$-language" as, respectively, "$\omega$-word" and "$\omega$-language." The disambiguation between an infinite language and an $\omega$-language, respectively denoted by $L^*$ and $L^\omega$, consists formally in the following result, which requires the specification $\Sigma^* = \bigcup_{i\geq 0}^{n<\omega} \Sigma^i$:

**Proposition 54.** *For $L^*$ and $L^\omega$ languages of finite and infinite words, respectively, we have*

$$L^* \cap L^\omega = \emptyset.$$

*Proof.* (Informal) For $\Sigma = \emptyset$, we have $L^* \subseteq \Sigma^* = \{\epsilon\}$, but $L^\omega \subseteq \Sigma^\omega = \emptyset$. Clearly, $\{\epsilon\} \cap \emptyset = \emptyset$. We now apply induction on $n > 0$: It is obvious that $(n + 1 = m) \neq (n + 1 = \omega)$ whenever $m < \omega$, and again we have $\Sigma^+ \cap \Sigma^\omega = \emptyset$. □

Let us now restrict the discussion to the case when, for a word $w$ s.t. $\ell(w) = k$ where $k \geq j$, we have

$$\underbrace{a_1 a_2 ... a_{j-1} a_j}_{w||_j} \quad a_{j+1} ... a_k$$

i.e. when the $j$-th letters of a word $w$ correspond to a (proper) prefix thereof.

**Definition 55.** The *prefix factorization* of a word $w \in \Sigma^+$ s.t. $\ell(w) = j$ consists of $j$ prefixes $w_1, w_2, ..., w_j$ s.t. $w_1 = a_1$, $w_2 = a_1 a_2$, etc., up to $w_j = a_1 a_2 ... a_j$. Then, the *prefix closure* of $w$ is defined as:

$$\widehat{w} = \left\{w_i \in \Sigma^+ \middle| \, i, j \in \mathbb{N} \text{ and } i \leq j\right\} = \{w_1, ..., w_{j-1}, w_j\}$$

s.t.

$$\{w_1\} \subset \{w_1, w_2\} \subset ... \subseteq \{w||_j\}.$$

Note that $w_0 = \epsilon$, a fact to take into consideration for a language over $\Sigma^*$, as we then have

$$\{w_0\} \subset \{w_0, w_1\} \subset ... \subseteq \{w||_j\}$$

for $j \geq 0$; then, $w \in \Sigma^*$ is factorized into $j + 1$ prefixes and the closure of the word $w$ s.t. $\ell(w) = j$ is given by $\widehat{w} = \{w_0, w_1, ..., w_j\}$.

**Definition 56.** A language $L \subseteq \Sigma^k$ is said to be *prefix-closed*, denoted by $\widehat{L}$, if we have the (possibly infinite) union

$$\widehat{L} = \bigcup \widehat{w}$$

for every $w \in L$.

*Remark* 57. Clearly, the language $L = \{\epsilon\}$ is prefix-closed, as we have $\widehat{\epsilon} = \{\epsilon\} = \widehat{L}$, but $L = \emptyset$ is not prefix-closed, as we have $\widehat{L} = \emptyset$.

**Example 58.** For instance, for the language $L = \{aa, aba, abba\}$ over $\{a, b\}^*$ the prefix closure of $L$ is the set:

$$\widehat{L} = \{\epsilon, a, aa, ab, aba, abb, abba\}$$

In effect, we have for the word $aa \in L$,

$$\underbrace{\underbrace{\underbrace{\epsilon}_{w_0} a}_{w_1} a}_{w_2}$$

and $w_0, w_1, w_2 \in \widehat{L}$.

The $\omega$-languages can also be prefix-closed. Below, in Remark 68, I account formally for this anti-intuitive result.

**Example 59.** The prefix-closure of the word $ab^\omega$ is $\widehat{ab^\omega} = \left\{a, ab, ab^2, ..., ab^i, ..., ab^\omega\right\}$. For the word $ab^\omega a$, we have the prefix-closure $\widehat{ab^\omega a} = \left\{a, ab, ab^2, ..., ab^\omega, ab^\omega a\right\}$, i.e. $\widehat{ab^\omega} \subset \widehat{ab^\omega a}$.

Languages are *generated* by formal grammars, in the sense that upon the application of syntactic rules on strings over a given alphabet $\Sigma$ words are *derived* that belong to a specific language $\Sigma^*$ or $\Sigma^\omega$. I next provide the definitions of these core terms.

**Definition 60.** A *formal grammar* is a 4-tuple $G = (V, T, S, P)$, where $V = \{A_1, ..., A_n\}$ is a set of variable symbols and $T = \{a_1, ..., a_m\}$ is a set of terminal symbols, with $(V \cap T) = \emptyset$, $S \in V$ is the start symbol, and $P = \{r_1, ..., r_k\}$ is a set of rewriting rules of the form

$$\underbrace{\alpha}_{\text{LHS}} \to \underbrace{\beta}_{\text{RHS}}$$

where "LHS" and "RHS" abbreviate "left-hand side" and "right-hand side," respectively, denoting that string $\alpha \in (V \cup T)^+$ can be rewritten as string $\beta \in (V \cup T)^*$ iff $\alpha$ has at least one variable symbol.

Rewriting rules are also called production rules or productions; hence the $P$. In order to distinguish clearly between variables and terminals uppercase and lowercase letters, respectively, are used. The first production rule $r_1$ is always $S \to \beta$. The language $L$ generated by a formal grammar is denoted by $L(G)$.

**Definition 61.** A *derivation* of a word $w \in L(G)$ from $S \in V_G$ is a finite sequence of steps each of which is the application of some production rule in $\{r_i\}_{i=1}^k \subseteq P_G$:

$$\mathscr{D}_w = S \overset{1}{\underset{r_1}{\Longrightarrow}} \alpha_1 \overset{2}{\underset{r_i}{\Longrightarrow}} \alpha_2 \overset{3}{\underset{r_i}{\Longrightarrow}} ...\alpha_n \overset{n}{\underset{r_i}{\Longrightarrow}} w$$

A derivation $\alpha \Longrightarrow \beta$ is said to be *leftmost* (*rightmost*), denoted by $\alpha \Longrightarrow_l \beta$ (respectively $\alpha \Longrightarrow_r \beta$), if at each step a production is applied to the leftmost (respectively rightmost) variable in $\alpha$. Accordingly, a grammar $G$ that generates leftmost (rightmost) derivations is called a *left-derivation* (*right-derivation*) *grammar*.

**Definition 62.** The sequence of the $k$ rules of a leftmost grammar applied in $n$ steps in a derivation $\mathscr{D}_w$ is called a *parse* and is defined as

$$\mathscr{P}_w = r_{1,1} r_{i,2} r_{i,3} ... r_{i,n}$$

where $r_{i,j}$, $1 \leq i \leq k$, denotes that rule $r_i$ is the $j$-th rule applied for $1 \leq j \leq n$.

Every leftmost (rightmost) grammar can be converted into a rightmost (leftmost, respectively) grammar, but here I shall focus on leftmost grammars (henceforth just "grammars"). A parse $\mathscr{P}_w$ is typically constructed by means of a tree $\mathcal{T}_w = (N, E)$ where $N$ is the set of nodes, labeled by variables in the case of the root $(S)$ or the interior nodes and by the terminals in $w$ if they are leaves, and $E$ is the set of edges joining two nodes. Clearly, every derivation $\mathscr{D}_w$ corresponds to a parse $\mathscr{P}_w$, and reciprocally. The above definition allows for the conception of an alphabet as $\Sigma = (V \cup T)$, as we have for a given string $\alpha \in \left(\Sigma^* = (V \cup T)^*\right)$, but we have $w \in (T^* = ((\Sigma - V) \cup \{\epsilon\}))$. That is to say that no string is a word if it contains at least one variable; this, in turn, entails that no LHS of a production rule may consist entirely of terminal symbols. This defines a *Chomsky grammar G* if (i) $\Sigma$ is finite and (ii) $L(G)$ is associated with a machine model $M$ s.t. $L(G) = L(M)$ in the sense that $L(G)$ is recognized by $M$, where by "recognition" it is understood that $M$ can completely process an input word one symbol at a time halting then in an accepting state. Table 1 provides all the information on Chomsky grammars required in this work. Note that this table schematizes more precisely the *extended* Chomsky hierarchy, i.e. the hierarchy of the languages generated by the four grammar types 0 through 3, known as Chomsky grammars, extended by the recursive languages, for which no grammar is (yet) known (cf. Fig. 1; see Table 1 for the nomenclature).[12] Note the following result about the Chomsky hierarchy:

**Proposition 63.** *The Chomsky hierarchy is a proper inclusion hierarchy:*

$$\mathscr{RGL} \subset \mathscr{CFL} \subset \mathscr{CSL} \subset \mathscr{REL}$$

*Proof.* Trivial. There are regular languages that are not context-free languages, there are context-free languages that are not context-sensitive languages, and there are context-sensitive languages that are not recursively enumerable languages. □

*Remark* 64. We also have

$$... \subset \mathscr{CSL} \subset \mathscr{RL} \subset \mathscr{REL}$$

if we consider the extended Chomsky hierarchy, which simply means what we call recursively enumerable languages are word sets whose complements are not recursively enumerable, i.e. semi-computable.

I can now define a language more precisely as follows:

**Definition 65.** Let $G$ be a formal grammar. The language $L = L^*$ generated by $G$ is defined as

$$L(G) = \left\{ w \in T^* | \, S \overset{*}{\underset{G}{\Longrightarrow}} w \right\}$$

where "$\Longrightarrow$" denotes a derivation step s.t. we have $S \overset{n}{\underset{G}{\Longrightarrow}} \alpha$ for the string $\alpha \in (V \cup T)^*$ in $n$ derivation steps, and "*" (Kleene star) denotes the reflexive and transitive closure of the relation $\underset{G}{\Longrightarrow} \subseteq (V \cup T)^+$.

[12] Recall that a recursively enumerable set is a semi-computable set and a recursive set is a computable set. In the (extended) Chomsky hierarchy, the recursion-based nomenclature has been kept for languages, despite a move to differentiate the terminology of recursion theory and computability theory, as noted in the Introduction.

| **Grammar** | $\alpha \to \beta$ | **Language Class & Prototypical Language** | **Recognizers** |
|---|---|---|---|
| *Type 0* Unrestricted | $\alpha \in (V \cup T)^{+}, \lvert V(\alpha) \rvert \geq 1$ <br> $\beta \in (V \cup T)^{*}$ | Recursively enumerable ($\mathscr{REL}$) <br> $L = \{(\mathcal{M}, w) \mid \mathcal{M} \text{ halts on } w\}$ | Turing machines (TMs) |
| | | Recursive ($\mathscr{RL}$) <br> $L = \left\{a^{2^n} \mid n \geq 0\right\}$ | Total TMs (Deciders) |
| *Type 1* Context-sensitive | $\alpha, \beta \in (V \cup T)^{+}, \lvert\beta\rvert \geq \lvert\alpha\rvert$ <br> $\alpha = \gamma A \delta$ <br> $\beta = \gamma X \delta$ <br> $\gamma, \delta \in (V \cup T)^{*}$ <br> $X \in (V \cup T)^{+}$ | Context-sensitive ($\mathscr{CSL}$) <br> $L = \{a^n b^n c^n \mid n > 0\}$ | Linear-bounded automata |
| *Type 2* Context-free | $\alpha \in V, \lvert\alpha\rvert = 1$ <br> $\beta \in (V \cup T)^{*}$ | Context-free ($\mathscr{CFL}$) <br> $L = \{a^n b^n \mid n > 0\}$ | Pushdown automata |
| *Type 3* Regular | $\alpha \in V, \lvert\alpha\rvert = 1$ <br> $\beta = Av$ or $vA$ <br> $A \in V^{*}, v \in T^{*}$ | Regular ($\mathscr{RGL}$) <br> $L = \{a^n \mid n > 0\}$ | Finite-state automata |

Table 1: The extended Chomsky hierarchy. (Adapted from Augusto, 2021.)

This definition applies directly to the Chomsky languages.

**Example 66.** Consider the formal grammar $G = (\{S, A, B\}, \{a, b\}, S, P)$ with

$$P = \left\{ \begin{array}{cc} (r_1) & S \rightarrow aA \\ (r_2) & A \rightarrow aAB \,|\, a \\ (r_3) & B \rightarrow b \,|\, \epsilon \end{array} \right\}.$$

This is a Type-1 Chomsky grammar. The language generated by this formal grammar is:

$$L(G) = \{a^n b^m |\, n \geq 2, m \geq 0\} =$$

$$= \{aa, aaa, aaaa, ..., aab, aabb, aabbb, ..., aaab, ...\}$$

The derivation of the word $a^3 b$ is:[13]

$$S \underset{r_1}{\Longrightarrow} aA \underset{r_2}{\Longrightarrow} aaAB \underset{r_2}{\Longrightarrow} aaaB \underset{r_3}{\Longrightarrow} aaab$$

The parse of this word is constructed by means of a tree as displayed in Figure 2.

This language is prefix-closed: For example, for the word $aaab$ we have the prefix closure:

$$\widehat{aaab} = \{a, aa, aaa, aaab\}$$

Note that $a \notin L(G)$, but it is a proper prefix of $aaab \in \widehat{L(G)}$.

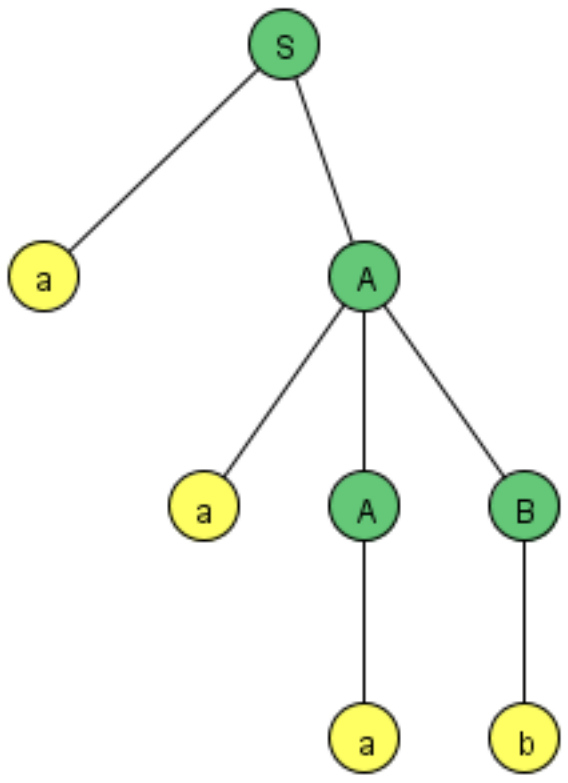

Figure 2: Parse tree $\mathcal{T}_{aaab}$.

We can extend Definition 65 above to the $\omega$-languages by specifying for $L = L^{\omega}$

$$L(G) = \left\{ w \in T^{+} |\, S \overset{\omega}{\underset{G}{\Longrightarrow}} w \right\}$$

where $G$ is an $\omega$-grammar.

[13] Note that I abbreviated the rules by using "|"; in fact, there are five rules in this grammar, but its simplicity allows for this abbreviation.

**Definition 67.** A formal grammar $G = (V, T, S, P)$ s.t. $P$ contains at least one rule of the form

$$\alpha \rightarrow \beta\gamma\delta$$

where $\alpha \in (V^+ \cup T^*)$, $\beta, \delta \in (V \cup T)^*$, and $\gamma \in V^\omega$, is called an *$\omega$-grammar* if $V^\omega = (V^+)^\omega$.

The notation $(V^+)^\omega$ denotes that one (at least) or more variables in $V$ is/are infinitely iterated. Let it be the case that the RHS of a rule is $bA^\omega c$ and we have $A \rightarrow a$; then, we have the partial derivation

$$bA^\omega c \Longrightarrow baA^\omega c \Longrightarrow ba^2 A^\omega c ... \overset{i}{\Longrightarrow} ba^{2+i} A^\omega c ... \overset{\omega}{\Longrightarrow} ba^\omega c$$

where $ba^\omega c$ is an $\omega$-word, but $a^\omega$ denotes an *infinite sequence* of $a$'s.[14]

*Remark* 68. The rule $A \rightarrow a$ is thus a continuous mapping $f_\omega : V^\omega \longrightarrow T^\omega$ that is in fact an extension of an unbounded sequential function $f_* : V^* \longrightarrow T^*$. In effect, consider that for $w_i^n$, denoting that a subword $w_i$ is iterated $n$ times, we have $w_i^\infty = \lim_{n \rightarrow \infty} = w_i^\omega$. This allows for the prefix closure of an $\omega$-language, inasmuch as we have for a given $\omega$-word $w$,

$$\widehat{w} = \left\{ w_1, ..., w_i^1, w_i^2, ..., w_i^\omega, ..., w_j \right\}$$

where $1 \leq i \leq j$, and $|\widehat{w}| = \infty$ but extensionally *finitely representable.*

**Example 69.** Consider the $\omega$-language $L = \{w \in \{a, b\}^\omega \mid w = ab^\omega\}$. The $\omega$-grammar $G = (\{S, A, B\}, \{a, b\}, S, P)$ where

$$P = \left\{ \begin{array}{cc} (r_1) & S \rightarrow AB^\omega \\ (r_2) & A \rightarrow a \\ (r_3) & B \rightarrow b \end{array} \right\}$$

generates this language. As seen above in Example 59, this $\omega$-language is prefix-closed. A derivation of $ab^\omega$ is:

$$S \underset{r_1}{\Longrightarrow} AB^\omega \underset{r_2}{\Longrightarrow} aB^\omega \underset{r_3}{\Longrightarrow} abB^\omega \underset{r_3}{\overset{i}{\Longrightarrow}} ab^{i+1} B^\omega \underset{r_3}{\overset{\omega}{\Longrightarrow}} ab^\omega$$

The parse tree $\mathcal{T}_{ab^\omega}$ is displayed in Figure 3, where the ellipsis denotes the infinite derivation $B \overset{\omega}{\Longrightarrow} b$.

The fact that, as pointed out above in Remark 68, the set $\widehat{w}$ for $w$ an $\omega$-word is extensionally finitely representable, allows for the recognition of $\omega$-languages by finite automata. The class of recognizers for the $\omega$-languages is the so-called $\omega$-automata, of which the Büchi automata are the *prima inter pares.* See Staiger (1997) for the case of the ultimately periodic $\omega$-languages, and Augusto (forthcoming) for the class of the $\omega$-regular languages, a special subclass of these languages.

### 3.1.2 3.1.2. Turing-decidability for languages

Recall from the remark on Definition 62 above that every word derivation $\mathscr{D}_w$ corresponds to a word parse $\mathscr{P}_w$, and reciprocally. However, it might not be easy to construct a parse tree for a given derivation. This is typically the case for the Type-0 Chomsky grammars.

[14] Of course, an $\omega$-word can coincide with an infinite sequence, say, $a^\omega$ or $(ba)^\omega$; in other words, an infinite sequence of terminal letters is any of a prefix, a suffix, or an infix. In the literature, $\omega$-words are typically restricted to words of the form $vz \in L^\omega$ where $v \in T^*$ and $z \in T^\omega$. See, e.g., Staiger (1997) for a formal discussion of these $\omega$-languages.

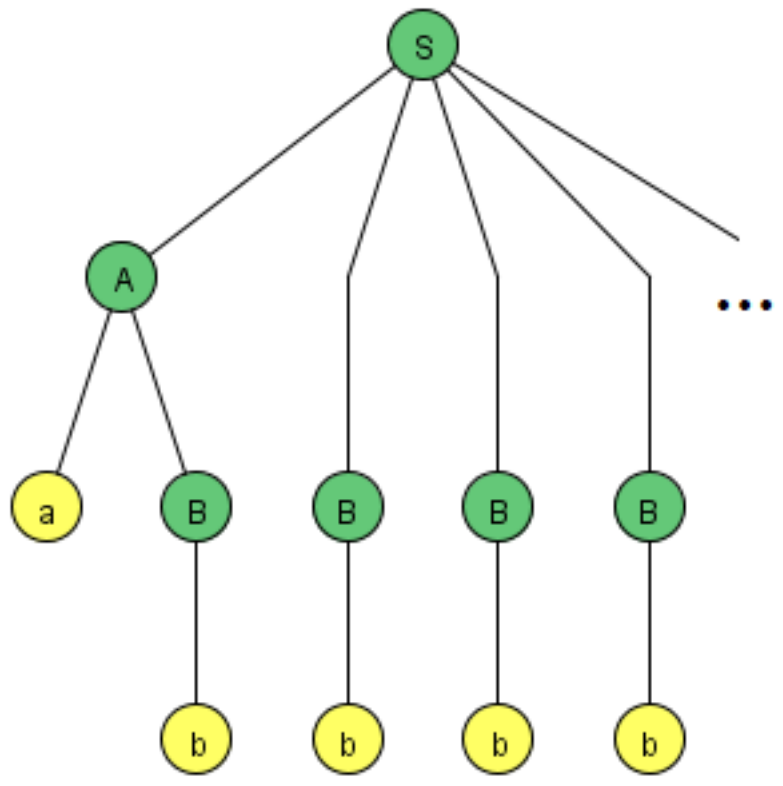

Figure 3: Parse tree $\mathcal{T}_{ab^\omega}$.

**Example 70.** Consider the formal grammar $G = (\{S, A, B\}, \{a, b\}, S, P)$ with

$$P = \left\{ \begin{array}{cc} (r_1) & S \to BA \\ (r_2) & A \to \epsilon \\ (r_3) & B \to aBb \\ (r_4) & aBb \to aa \\ (r_5) & bA \to bbbA \end{array} \right\}.$$

This is a Type-0 Chomsky grammar. The language generated by this grammar is:

$$L(G) = \{a^n b^m | \, n \geq 2, m = n - 2 \text{ or } m = n + 2k, k \geq 0\}$$

We can derive the word $a^2 b^4 \in L(G)$ as follows:

$$S \underset{r_1}{\Longrightarrow} BA \underset{r_3}{\Longrightarrow} aBbA \underset{r_5}{\Longrightarrow} aBbbbA \underset{r_3}{\Longrightarrow} aBbbbbbA$$

$$\underset{r_4}{\Longrightarrow} aabbbbA \underset{r_2}{\Longrightarrow} aabbbb$$

The parse tree for this word is shown in Figure 4.

**Definition 71.** Let $L \subseteq \Sigma^*$ be a language. The *decision problem* for $L$, denoted by $DP_L$, consists in finding a Turing machine $\mathcal{M}$ capable of answering the question $w \overset{?}{\in} L$, known as the *membership problem.*[15] Then, $L$ is said to be *Turing-decidable* if there is a Turing machine $\mathcal{M}$ that computes

$$DP_L(w) = \begin{cases} \text{Yes} & \text{if } w \in L \\ \text{No} & \text{if } w \notin L \end{cases}$$

s.t. we have the sets

$$L = \{w \in \Sigma^* | \, \mathcal{M} \text{ halts on input } w \text{ in state } q_a\}$$

[15] The decision problem is also known in the literature as the *Entscheidungsproblem*, because of its first formulation in German (for logical formulas) by the mathematician D. Hilbert. Cf. Hilbert & Ackermann (1928).

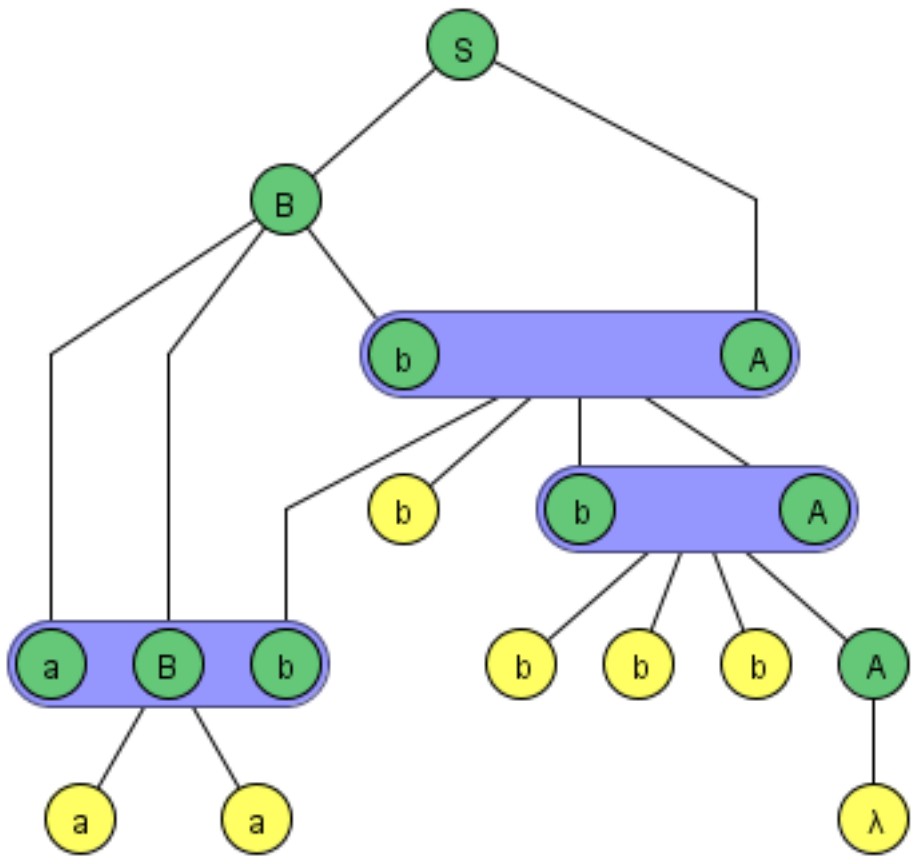

Figure 4: Parse tree $\mathcal{T}_{a^2b^4}$.

where $q_a$ denotes an accepting state, and

$$\overline{L} = \{w \in \Sigma^* | \mathcal{M} \text{ halts on input } w \text{ in state } q_r\}$$

where $q_r$ denotes a rejecting state.

*Remark* 72. This is to say that a Turing machine $\mathcal{M}$ decides a language $L$ iff it halts on every input word $w$, whether in an accepting or a rejecting state. If $\mathcal{M}$ halts in an accepting state when $w \in L$ for a given language $L$, but fails to halt in a rejecting state whenever $w \notin L$, then we say that $L$ is a *semi-decidable* language. If $\mathcal{M}$ never halts on an input word $w$, then $L$ s.t. $w \in L$ is *undecidable*.

Recall Remark 64 above. We have the following result:

**Theorem 73.** *Every Chomsky grammar of Types 3 through 1 generates decidable languages.*

*Proof.* (Informal) In the extended Chomsky hierarchy, we have the strict inclusion relation:

$$\mathscr{RGL} \subset \mathscr{CFL} \subset \mathscr{CSL} \subset \mathscr{RL}$$

□

**Definition 74.** Note that I am employing here a Turing machine as a 7-tuple $\mathcal{M} = (Q, \Gamma, \#, \Sigma, q_0, H, \delta)$ where $Q$ is a finite set of states, $\Gamma$ is the tape alphabet, $\# \in \Gamma$ is the blank-cell symbol, $\Sigma \subseteq (\Gamma - \{\#\})$ is the input alphabet, $q_0 \in (Q - H)$ is the initial state, $H = \{q_a, q_r\}$ for $H \subseteq Q$ is the set of halting states, and $\delta : ((Q - H) \times \Gamma) \longrightarrow (Q \times \Gamma \times \{L, R\})$ is the transition function with $L$, $R$ denoting left and right motion, respectively. A configuration for $\mathcal{M}$ is a pair $C_{i \in \mathbb{N}} = (q, uav)$ indicating that $\mathcal{M}$ is in state $q \in Q$ ($q_0 \in Q$ for the initial configuration $C_0$) with the current tape string $uav$ and reading the symbol $a$. Configuration $C_i$ yields configuration $C_{i+1} = (p, xby)$ in one step or move, written $C_i \vdash_{\mathcal{M}} C_{i+1}$, iff the transition $\delta(q, a)$ changes configuration $C_i$ to configuration $C_{i+1}$, i.e. iff we have $\delta(q, a) = (p, b, \triangledown)$ where $p \in Q$, $b \in \Sigma$ is the (new) symbol on the tape, and $\triangledown \in \{L, R\}$. A computation for $\mathcal{M}$ is a finite sequence of configurations $C_0, C_1, ..., C_n$ s.t. *Conf* $(\mathcal{M}, w) = \{C_i\}_{i=0}^n$ is the set of all the configurations of $\mathcal{M}$ when computing a given input word $w$.

For simplicity, unless otherwise stated a Turing machine $\mathcal{M}$ is deterministic, i.e. $|\delta(q,a)| = 1$ for every $q \in Q$ and $a \in \Gamma$. Clearly, the Turing machine $\mathcal{M}$ in the definition above effectively calculates the characteristic function $\chi_L(w)$ (cf. Def. 6 above) if we make "1" stand for "$q_a$" and "0" do so for "$q_r$." In other words, a language $L$ is Turing-decidable iff the set $L$ is *Turing-computable*, a property denoted by $L(\mathcal{M})$. An alternative formal way to formulate this is by the notion of *enumeration*, which was already discussed above, but before I address that subject let us agree—with Post (1944)—that "recursive" is the technical term corresponding to the more intuitive "effectively calculable" when speaking of functions. Turing (1936) precisely defined Church's *effective calculability* as *mechanical computability*; in fact, he showed that all computable numbers are effectively calculable by means of the Turing machine. This, now known as *Turing's thesis*, allows us to exchange the term "recursive" by the term "Turing-computable" in view of the short discussion in the Introduction, and because no function is computable if it is not computable by a Turing machine we shall henceforth satisfy ourselves with the abbreviation "computable."

**Lemma 75.** *No configuration $C_{i+1} \in Conf(\mathcal{M},w)$ of a Turing machine $\mathcal{M}$ is a proper prefix of a configuration $C_i \in Conf(\mathcal{M},w)$.*

*Proof.* For every word $w$ s.t. $\ell(w_{i+1}) > \ell(w_i)$, we have

$$\underbrace{w_1}_{C_1}, \underbrace{w_2}_{C_2}, ..., \underbrace{w_n}_{C_n}$$

where every $C_j$, $1 \leq j \leq n$, is a unique configuration of $\mathcal{M}$. □

**Definition 76.** Let $\mathcal{M}$ be a Turing machine with $n$ tapes, $n \in \mathbb{Z}^+$, s.t. the content on tape $i$, $1 < i \leq n$, is a word ${}^i w$, denoting the word $w$ on the $i$-th tape. (We may consider that every ${}^i w$ is a word that has been accepted by $\mathcal{M}$ so that $L(\mathcal{M}) = \left\{{}^i w\right\}_{i=2}^{n}$ is the language recognized by $\mathcal{M}$.) On every $i$-th tape the content is ${}^i w = a_1 a_2 ... a_k \# ...$, $k \geq 1$, with symbol $a_1$ at the leftmost cell of the tape and s.t. $a_k$ is immediately followed by an infinite number of empty cells. Let tape 1 of this Turing machine be a one-way write-only tape s.t. at the end of the computation the content of tape 1 is ${}^i_1 w \#^i_2 w \# ... \#^i_{n-1} w \# ...$ where all $i$'s are different and ${}^i_l w$ denotes that ${}^i w$ is the $l$-th word on the sequence. This $\mathcal{M}$ *enumerates* $L$. $\mathcal{M}$ enumerates all the strings of $L$ in *canonical order* if shorter strings precede longer strings and strings of the same length are alphabetically ordered.

Clearly, a Turing machine enumerates a language $L(\mathcal{M})$ in canonical order iff it computes $\widehat{L(\mathcal{M})}$, i.e. iff (a) for every word ${}^i w \in \Sigma^k$ it computes $\left\{{}^i w_1, {}^i w_2, ..., {}^i w_k\right\} = \widehat{{}^i w}$ and (b) in such a way that $\ell\left({}^i_{l+1} w\right) \geq \ell\left({}^i_l w\right)$ and $\left({}^i_l a_1 \preceq {}^i_{l+1} a_1\right), ..., \left({}^i_l a_k \preceq {}^i_{l+1} a_k\right)$ where ${}^i_l a_j$ denotes the $j$-th letter of the $j$-th prefix of the $l$-th word.

**Theorem 77.** *For every language $L \subseteq \Sigma^*$, $L$ is semi-Turing-decidable iff there is a Turing machine $\mathcal{M}$ that enumerates $L$, and $L$ is Turing-decidable iff there is a Turing machine $\mathcal{M}$ that enumerates the elements of $L$ in canonical order.*

*Proof.* (Informal) If $\mathcal{M}$ simply enumerates $L$, then $L$ is clearly a semi-computable set, but if $\mathcal{M}$ enumerates $L$ in canonical order, then both $L$ and $\overline{L}$ must be semi-computable by the Complementation Theorem (cf. Theorem 13). □

**Corollary 78.** *A language $L$ is Turing-decidable iff it is a prefix-closed language $\widehat{L(\mathcal{M})}$.*

*Proof.* (Informal) By Lemma 75, every Turing machine can compute $\widehat{L(\mathcal{M})}$ for a given language $L(\mathcal{M})$. If a language $L$ is not prefix-closed, its words cannot be enumerated by a Turing machine $\mathcal{M}$ in canonical order. In particular, for every word $w \in L \subseteq \Sigma^k$, there is a prefix $w_k$ s.t. $\mathcal{M}$ computes $w_{k-1}$, $\ell(w_{k-1}) < \ell(w_k)$, but $\mathcal{M}$ fails to compute $w_k$ s.t. $\ell(w_k) \leq \ell(w)$. □

The Turing machine is a wholly abstract machine; in effect, it is equipped with one or more infinite tapes. In practice, we work with *parsers*, also called syntax analyzers. An interesting aspect of parsers is that they emulate the workings of a Turing machine almost to perfection: Given some grammar $G$ with alphabet $\Sigma$, there might be input words in $\Sigma^+$ on which they do not stop. This means that a parser is as good a Turing machine as we can get an implementation thereof. For this reason, I henceforth write simply "decidable," instead of "Turing-decidable" with respect to a language $L(G)$.

**Example 79.** Consider again the language $L(G)$ generated by the grammar $G$ of Example 70 above. The parser implemented by the software jflap accepts all words in $L(G)$ but it does not halt on any word in $\{a,b\}^+$ that does not belong to this language; more precisely, this parser is unable to construct the parse tree of any such word, entering into an infinite node generation. Hence, this Type-0 grammar generates a semi-decidable language.

## 3.2 Productive grammars and productive languages

As seen above, a set $A \subseteq \mathbb{N}$ may be only semi-computable or even computable, the latter case depending on whether both $A$ and its complement $\overline{A}$ are semi-computable sets. Languages are just sets of words; in the jargon of formal languages here adopted, a language $L$ is called decidable if its is a computable set and semi-decidable if it is only semi-computable. This terminology was set by employing Turing machines. In Section 2.2 above, it was shown that productive sets are incomputable. Then, if we construct languages that behave like productive sets, we have languages that are not only undecidable, but actually essentially undecidable languages; in this case, there is no Turing machine that can decide for each word $w \in L$, where $L$ is defined intensionally, whether $w$ is one of the words in $L$.

**Definition 80.** Let $G$ be a grammar. If $G$ has at least one production rule on strings of $n$ letters that mimics a productive function on $n \in \mathbb{N}$, called a *productive rule*, then $G$ is called a *productive grammar*. A language $L = L(G)$, where $G$ is a productive grammar, is called a *productive language*.

**Example 81.** Consider the grammar $G = (\{S, A, B\}, \{a, b\}, S, P)$ with:

$$P = \left\{ \begin{array}{c} (r_1) \quad S \to aA \\ (r_2) \quad aA \to_l aaAB \\ (r_3) \quad B \to b \end{array} \right\}$$

The subscript "$l$" in rule $r_2$ on "$\to$" prescribes that this rule is compulsorily a leftmost-derivation rule. We have $L(G) = \{a^n b \mid n \geq 1\}$, but it is evident that $a^n b \notin \widehat{L(G)}$ for any $n \in \omega$ s.t. $a^n$ is a prefix. This is so because there is a single infinite derivation

$$S \underbrace{\Longrightarrow}_{r_1} aA \underbrace{\Longrightarrow_l}_{r_2} aaAB \overbrace{\underbrace{\Longrightarrow_l}_{r_2} \ldots \underbrace{\Longrightarrow_l}_{r_2}}^{\omega} a^\omega aAB$$

s.t. $r_3$ is never applied and there is no terminal word $a^n b$. Clearly, no Turing machine $\mathcal{M}$ can enumerate this language, let alone enumerate it in canonical order, as for every $a^n$ it is the case that $a^n b \notin \widehat{L(G)}$, but $\mathcal{M}$ halts neither in an accepting nor in a rejecting state. In effect, $\mathcal{M}$ has no means to determine, for any $n \in \omega$, that $a^n$ is a proper prefix of $a^n b$. Thus, $L(G)$ is essentially undecidable.

Note with respect to this language that, were $G$ a Chomsky grammar (of Type 1 or 0), we would have $L(G) = \emptyset$, because rule $r_2$ hinders the derivation of any string $w \in T^+$.

**Theorem 82.** *Productive languages are outside the Chomsky hierarchy.*

*Proof.* (Informal) The set $P$ in a Chomsky grammar $G = (V, T, S, P)$ is a set of production rules. Whenever there is a rule $r \in P$ s.t. $r$ prevents the rightmost or leftmost derivation of a word, then we have $L(G) = \emptyset$. If $P \supseteq \{r\}$ in a grammar $G = (V, T, S, P)$ s.t. $r$ is a productive rule, then $L(G) \neq \emptyset$, and $L(G)$ is a productive language. □

A similar theorem can be stated for the $\omega$-languages and the proof is based on the fact that no productive language has a rule of type $\alpha \to \beta\gamma\delta$ specified in Definition 67 above. The distinction between Chomsky or $\omega$- languages and productive languages falls thus on the productive rules. I next analyze formally these rules.

**Definition 83.** Let $w_n$ be a prefix of a word $w = w_{n+1}$. Clearly, $w_n$ is a proper prefix of $w$, and because $w_n$ is only one letter away from $w$ let us call it the *terminal proper prefix* of $w$. For every $n$, we define the function $t : \mathbb{N} \longrightarrow \Sigma^+$ s.t. $t(n) = w_{n+1}$, i.e. $t(n)$ sends the terminal proper prefix of $w$ to $w$ itself. I shall call $t$ the *word terminating function*.

Clearly, $t(n)$ is total and it is an increasing function. Hence, $t$ is a recursive function even if $|Dom(t)| = \omega$. Then, the set

$$L = \{w \mid \forall n \in \mathbb{N}, t(n) \downarrow \text{ and } t(n) = w\} =$$
$$= \{t(0), t(1), t(2), ...\} = Range(t)$$

is computable even if its domain is infinite, and because $L$ is a language, i.e. a word set, we say that $L$ is decidable.

**Example 84.** Consider the language $L = \left\{a^n b \in \{a, b\}^{n+1} \mid n \geq 1\right\}$. We have

$$t(1) = ab$$
$$t(2) = aab$$
$$\vdots$$
$$t(n) = a^n b$$
$$\vdots$$

It is easy to see that a Turing machine can enumerate the words of $L$ in canonical order. We thus have $\widehat{L(\mathcal{M})}$ and $L$ is a decidable language. This language could be generated by a regular grammar, but because we are interested in Turing-decidability let us construct a Type-0 grammar for it. This can be the grammar $G = (\{S, A, B\}, \{a, b\}, S, P)$ with:

$$P = \left\{ \begin{array}{cc} (r_1) & S \to aAB \\ (r_2) & aA \to aaA \\ (r_3) & AB \to b \end{array} \right\}$$

It is also easy to see that $\overline{L(G)}$ is also semi-decidable, so the language generated by this grammar is actually decidable.

**Definition 85.** Let us now set $W_n = \{w \in \Sigma^n | \, t(n) = w_{n+1}\}$, $W_n \subseteq L$ for a given language $L$. If we now make

$$W_n \subseteq L \Rightarrow \begin{cases} (1) & t(n) \\ (2) & t(n) \in (L - W_n) \end{cases}$$

then $t(n)$ is a productive function for $L$ and $L$ is a productive language.

In effect, $W_n$ is an infinite semi-computable set, but $L \neq W_n$ for any subset $W_n$ of $L$. This means that $L$ is not a prefix-closed language, being thus an undecidable language. In the formalism of productive sets (cf. Section 2.2 above), we have the set $\Pi(L, t)$, i.e. the productive center of $L$ relative to the productive function $t(n)$. Then we have

$$W_n \subseteq L \Rightarrow \begin{cases} (1) & s(n) \downarrow \\ (2) & W_{s(n)} \subseteq (L - W_n) \\ (3) & \left|W_{s(n)}\right| = \omega \end{cases}$$

for a recursive function $s(n)$ s.t. $W_{s(n)} = W_n \cup \{w\}$ by defining

$$\varphi_{s(n)}(w) = \begin{cases} 1 & \text{if } w \in W_n \text{ or } w = t(n) \\ \uparrow & \text{otherwise} \end{cases} .$$

We can then apply the results constituted by Proposition 41 through Corollary 46 to a productive language $L$.

Consider again the language of Example 84, but now generated by the grammar $G$ of Example 81 above. It is easy to see that $L(G)$ is undecidable, because $\widehat{w} \not\supseteq \{w\}$ for any word $w \in L(G)$ we have it that $w \notin W_n$. But $\widehat{L(G)}$ contains an infinite semi-computable subset:

$$\widehat{L(G)} \supset \{a, aa, aaa, ..., a^{\omega}\}$$

Additionally, $\overline{L(G)}$ is semi-computable. Thus, $G$ is a productive grammar and $L(G)$ is the productive language generated by $G$.

# 4 Conclusions

In the Introduction to this article, I justified the study of productive languages by the need to reduce our ignorance with respect to the infinitely large class of undecidable languages. There are in effect infinitely many undecidable languages, and uncountably infinitely many for that matter, a fact that finds its theoretical support in the following results in recursion or computability theory.

**Lemma 86.** *There are exactly countably infinitely many p.r. functions and there are exactly countably infinitely many recursive functions.*

I now convert this result into the framework of Turing machines:

**Theorem 87.** *Let $M$ be the set of all Turing machines. Then, $|M| = |\omega|$, i.e. there are countably infinitely many Turing machines.*

*Proof.* (Informal) There are as many Turing machines as (partial) recursive functions. □

Let us now denote an uncountable infinity by $2^{|\omega|}$and remark that $|\omega| \leq 2^{|\omega|}$, a result that follows from Cantor's theorem, which equates with stating that the power set $2^{\omega}$ of $\omega$ is uncountable. Then, it follows that there are languages that are not Turing-recognizable—a result that I have proven above constructively via the productive languages. Formally, we have:

**Corollary 88.** *There are uncountably infinitely many languages that are undecidable.*

*Proof.* Let $\mathscr{PL}$ denote the class of all productive languages. I have shown above constructively that, alone for $\mathscr{PL}$, we have already $|\mathscr{PL}| \geq |\omega|$. □